UseRawInputEncoding
\documentclass[prl,aps,onecolumn,showpacs]{revtex4}

\usepackage{psfrag}
\usepackage{graphicx}
\usepackage{graphics}
\usepackage{bm}
\usepackage{color}
\usepackage{centernot}

\newcommand{\be}{\begin{equation}}
\newcommand{\ee}{\end{equation}}
\newcommand{\ba}{\begin{eqnarray}}
\newcommand{\ea}{\end{eqnarray}}

\newcommand{\dcom}[1]{}
\newcommand{\dnote}[1]{}

\newcommand{\gsim}{\raise.3ex\hbox{$>$\kern-.75em\lower1ex\hbox{$\sim$}}}
\newcommand{\lsim}{\raise.3ex\hbox{$<$\kern-.75em\lower1ex\hbox{$\sim$}}}

\begin{document}

\renewcommand{\thefootnote}{\fnsymbol{footnote}}


\renewcommand{\thefootnote}{\arabic{footnote}}
\setcounter{footnote}{0} \typeout{--- Main Text Start ---}

\title{ 

Non-singular Bounce from a Phase transition}
\author{ C.~ C. Wong}
\affiliation{Department of Electrical and Electronic Engineering, University
of Hong Kong. H.K.}

\date{\today}
\begin{abstract}
The single field Ekpyrosis of Khoury and Steinhardt \cite{khoury}-\cite{khoury1} admits a Jordan frame scalar field model with a double well potential. After the end of the ekpyrotic phase, as energy density increases we find that a phase transition scenario to a symmetric phase at the conformal coupling fixed point leads to a non-singular bounce in Einstein frame. This scenario could admit a simple prescription for the small value of dark energy.
\end{abstract}

\pacs{??}

\maketitle
\section{Introduction}

The density fluctuations in the observed Cosmic Microwave Background \cite{ade1}-\cite{ade2} requires a nearly scale invariant curvature power spectrum within an early Hubble radius, see \cite{brandenberger}. The current paradigm to provide these fluctuations, as well as a flat, homogeneous and isotropic universe, is inflation [4]. Inflation can be realised using effective models of canonical quantum scalar fields with a plateau potential in Einstein frame \cite{brandenberger}-\cite{martin}.  From a similar scalar field plateau potential, one can also obtain nearly invariant curvature perturbations in slowly evolving (contracting or expanding) solutions, \cite{piao}-\cite{lyth}. For the inflation solution, conceptual challenges such as super-Planckian problem, multiverse problem aside \cite{brandenberger}-\cite{steinhardt}, the isotropic power loss of $5\sim10\%$ in the small multipole $l< 40$ range, suppression of quadruple and dipolar asymmetry \cite{ade2}, are notable concerns. Some attempts to address these issues involve pre-"slow-roll inflation" physics such as a constant roll phase \cite{starobinsky}, a fast roll phase \cite{linde}, a "slowly expanding" phase or a "slowly contracting" phase \cite{piao}. The slowly contracting phase we focus on in this work is the "Ekpyrotic" phase in \cite{khoury}-\cite{khoury1}, which requires a bounce to connect the contracting phase to the subsequent expanding phase, whether it be a radiation dominant phase or an inflation phase. A bounce can be non-singular if it occurs at energy scales sufficiently lower than the Planck scale. 
\\\\
To obtain a bounce mechanism presents a major challenge. After the Ekpyrotic phase, the Einstein frame Hubble constant is $H_E<0$ (and $\dot{H}_E<0$) and the scale factor of the universe is in a slowly contracting phase. For an effective bounce to occur, one requires a phase where $\dot{H}_E>0$ and after the bounce a subsequent field roll in a non negative potential $V\geq 0$, see \cite{xue}. In the scalar field kinetic energy dominant epoch, with a shrinking scale factor $a$, the field kinetic energy has the largest power $\epsilon$ in $a^{-2\epsilon}$ amongst cosmological fluids, spatial curvature and static field potential. This slowly contracting phase is shown to be an effective smoothing mechanism \cite{kist} and addresses the gravitational entropy problem \cite{steinhardt2021}. A contracting universe will keep contracting unless some "additional" conditions exist that leads to  $\dot{H}_E>0$. One early approach is by allowing, for a short time before the bounce, the kinetic or gradient term to take the wrong sign with its cocomitant violation of Null Energy Conditon (NEC) \cite{ljjas}. More careful analysis \cite{vikman} shows that the cosmological perturbations along the transition between dark energy field and phantom field is unstable. Some progress is found in a class of models with Kinetic Gravity Braiding \cite{vikman2} and Galileons \cite{ljjas1}-\cite{ljjas2}, though challenge remains \cite{vikman3}. Other notable solution involves higher derivatives operator \cite{piao2019}.
Attempts to obtain a bounce from canonical scalar fields include using a generic positive curvature \cite{matsui} and adding string inspired negative energy density term \cite{brandenberger2}. In Jordan frame, one can implement a bounce without Null Energy Condition (NEC) violation, \cite{li}-\cite{ljjas3}. Yet these solutions are shown to be dual to inflation in Einstein frame. A Jordan frame model with a conformal coupling involving the Higgs field is also considered in \cite{bars}-\cite{bars2}, in which the Einstein frame scalar field plateau potential is described as a double well potential in Jordan frame. A phase transition is effected when the quartic coupling of the scalar field passes between a positive value region and a negative value region. In this work, we consider a Jordan frame double well potential which goes through a phase transition at conformal coupling.
\\\\
As we consider a bounce after the end of Ekpyrotic phase in \cite{khoury}-\cite{khoury1}, we recall some of their results for completeness. Here one considers a canoncial scalar field with a lifted plateau potential in Einstein frame
\begin{equation}
V=V_1-V_0e^{-c\varphi/M_{pl} },
\label{v1}
\end{equation}
where $M_{pl}$ is the Planck mass, $V=-V_0e^{-c\varphi/M_{pl}}$ is the negative potential and $V_1$, usually taken as $V_1=V_0$, is the lift to take the overall potential to positive density at large $\varphi$. This potential provides an effective model for Inflation \cite{martin}. However, the same model also admits two other solution branches, namely adiabatic Ekpyrosis and a slow expansion followed by a slow contraction \cite{lyth}.
\\\\
Ekpyrosis scenario \cite{turok}-\cite{steinhardt2}, where a period of ultra slow scale factor contraction, with equation of state parameter $\epsilon \gg 1$, can be shown to produce scale invariant power spectrum fluctuations. For potential in Eq.(\ref{v1}),  we take $M_{pl}=1$, with scale factor $a(t)$ and Einstein frame Hubble constant $H_E$, where the cosmic time $t$ is negative and increasing toward $0_{-}$. At large c,  the field has a tachyonic roll given by the equation
\be 
\ddot{\varphi} +3H_E\dot{\varphi}+V_{,\varphi}=0.
\label{fieldeqn}
\ee 
Given that $-t_{ek-beg}$ and $-t_{ek-end}$ are the points where Ekpyrotic phase begins and ends, Eq.(\ref{fieldeqn}) has the following scaling solution 
\be
a(t) = \bigg(\frac{-t}{-t_{ek-end}}\bigg) ^{1/\epsilon};\:\:\:  H_E=\frac{2}{c^2t},\:\:\:\: \dot{H}_E=-\frac{2}{c^2t^2};\:\:\:\epsilon=\frac{3}{2}(1+\omega) =-\frac{\dot{H}_E}{H_E^2} = \frac{c^2}{2}
\label{ekpysol}
\ee
\be
\varphi=\frac{2}{c}\ln \bigg(-\sqrt{\frac{V_0}{2(1-6/c^2)}} ct\bigg).
\ee
Given the density $\rho$ and pressure $P$ of the cosmological fluid, the Einstein frame Hubble constant is given by
\be
3H_E^2=\rho= \frac{1}{2}\dot{\varphi}^2+V(\varphi);\:\;\: P=\frac{1}{2}\dot{\varphi}^2-V(\varphi),
\label{eos}
\ee
with $\rho=a^{-2\epsilon}$ and at nearly constant $\epsilon$, we obtain 
\be
6H_E\dot{H}_E=\dot{\rho}=-3H_E(\rho+P);\:\;\;\;\dot{H}_E=-\frac{1}{2}\dot{\varphi}^2=-\frac{1}{2}(\rho+P).
\label{nec}
\ee
where the Null Energy condition (NEC) is $\rho+P\geq0$. The Ekpyrotic phase describes a contracting scale factor with $H_E<0$ and $\dot{H}_E<0$. However, it is argued that the curvature spectrum of this canonical scalar field is necessarily blue-tilted \cite{creminelli0}
\\\\
By extrapolating the ekpyrotic scaling solution to much earlier time, the authors of \cite{khoury}-\cite{khoury1} find a "Pre-ekpyrotic " phase in which a scale invariant power spectrum can be made red-tilted.
\\\\
Specifically, by integrating $\dot{H}_E=-\frac{2}{c^2t^2}$ one obtains 
\be
H_E= \frac{2}{c^2t} -\sqrt{\frac{V_1}{3}}; 
\label{HE}
\ee
where the equation of state paramter $\epsilon=\frac{3}{2}(1+\frac{p}{\rho})$ is given by
\be
\epsilon =\frac{6}{c^2V_0} \frac{1}{(t+t_{ek-beg})^2}\sim \frac{1}{t^2}
\ee
for larger $|t|\gg |t_{ek-beg}|$ where $t_{ek-beg}$ is the time at which the pre-Ekpyrotic phase ends and the Ekpyrotic phase begins. The equation of state parameter $\epsilon$ changes from a small value at $t^2\gg1$ towards a large constant $\frac{c^2}{2}$ as time grows towards $t_{ek-beg}$.
\\\\
In this large $t^2$ "Pre-Ekpyrotic" phase, from Eq.(\ref{HE}) the Hubble radius is large and the small momentum curvature perturbations grow outside of a curvature scale (a reminiscence of the correlation length) that is smaller than the Hubble radius. These scale invariant perturbations (due to the small $\epsilon$) which is red-tilted stay inside the Hubble radius and can only exit the Hubble radius during a subsequent Ekpyrotic (contracting Hubble radius) phase. During the ekpyrotic contraction, the red tilt spectrum will exit the Hubble radius first to be followed by the blue tilt "ekpyrotic" spectrum. 
\\\\
Cosmological observations of non-gaussianity and a small red-tilt  requires that $c(\varphi)$ in Eq.(\ref{v1}) to be very large ($\sim 10^{28})$ \cite{khoury} in both the "pre-Ekpyrotic" phase and at the beginning of the Ekpyrotic phase. Also $c^{'}(\varphi) >0$ (that is $c$ decreases as $\varphi$ decreases) is required to reduce $c(\varphi)$ to $O(1)$ where the Ekpyrotic phase ends.
After $c(\varphi)$ takes $O(1)$ value, we have $\dot{H}_E =-\frac{1}{2} \dot{\varphi}^2$. To match observations that the Hubble radius is expanding at late time. $\dot{H}_E$ needs to change from a negative value to positive value ($\dot{H}_E>0$). Typically the connection between the contracting phase to expanding phase (the bounce) is described schematically as $a\rightarrow a$, $\dot{a}\rightarrow -\dot{a}$, $\varphi \rightarrow \varphi$, $\dot{\varphi}\rightarrow \dot{\varphi}$.
\\\\
In this work, by noting that the Einstein frame canonical scalar field with a potential in the form of Eq.(\ref{v1}) admits a corresponding Jordan frame scalar field with a double well potential, we consider that the system is in a broken symmetric phase. As energy density increases we study the scenario that this Jordan frame scalar field goes through a phase transition where the non-minimal coupling bounces off its conformal value with zero mass. We find that $\dot{H}_E>0$ appears naturally in the broken symmetric phase leading to a bounce into the symmetric phase. In the next section, we recapitulate the correspondence of Eq.(\ref{v1}) in Jordan frame.  We consider the scaling region in the broken symmetric and the symmetric phase, the bounce and the subsequent field roll in the effective potential in the symmetric phase. We consider the value of the cosmological constant in the Einstein frame in the symmetric phase and briefly consider the scenario of a phase transition back to the broken symmetric phase.  We end with a conclusion and discussion.
\subsection{The Jordan frame double well potential}
Starting with a Jordan frame action
\begin{equation}
S_J=\int d^4x \sqrt{-g_J} \bigg( \frac{1}{2}F(\phi) R_J +\frac{K}{2}g_J^{\mu\nu}\partial^{\mu}\phi\partial_{\mu} \phi -V_J(\phi^2)\bigg),
\label{s1}
\end{equation} 
for $K\leq 6$ (where $K=6$ is the conformal coupling value), where the positive kinetic energy in Jordan frame corresponds to $K>0$.
The Einstein frame action is given by (see for example \cite{darabi}
\begin{equation}
S_E=\int d^4x \sqrt{-g_E} \bigg(\frac{1}{2} M_{pl}^2R_E -M_{pl}^2\frac{1}{2}\partial^{\mu}\varphi\partial_{\mu}\varphi -M^4_{pl}\frac{V_J(\phi)}{F^2}\bigg)
\label{se}
\end{equation} 
$M_{pl}$ terms will be useful in a specific consideration later. Both $\phi$ and $\varphi$ fields are in units of $M_{pl}$ and we will take $M_{pl}=1$ for most of our discussion. Based on the usual assumption that $\dot{c}\ll c$
\begin{equation}
F=\phi^2=e^{c\varphi};\:\:\: \frac{\phi_{,\mu}}{\phi} =\frac{c}{2} \varphi_{,\mu}
\label{F}
\end{equation}
here for $\phi^2>1$ $\varphi>0$ and $\phi^2<1$, we have $\varphi<0$.
The Einstein frame canonical field is related to Jordan frame field by
\begin{equation}
\varphi_{,\mu}=\phi_{,\mu}\sqrt{\frac{3}{2}\bigg(\frac{F_{,\phi}}{F}\bigg)^2 -\frac{K}{F}}  = \bigg(\sqrt{6-K} \bigg)\frac{\phi_{,\mu}}{\phi}\:\:\:.
\label{phivarphi}
\end{equation}
Comparing Eq.(\ref{F}) with Eq.(\ref{phivarphi}). here $"c"$ is no longer a free parameter but is related to the non-conformal coupling $"K"$ by the equation
\be
c=\frac{2}{\sqrt{6-K}}.
\label{ck}
\ee
such that for near conformal coupling value $K\lsim 6$, the parameter $c$ takes on a very large value. 
\\\\
In Scalar-Tensor theory modelling of cosmological evolution from high energy density, as energy density $\rho$ reduces,  $K(\rho) <6 $ is usually taken as a fine-tunable parameter, while the parameters in the potential $V_J$ is either taken to be fixed or to vary with $K$, see for example \cite{wetterich2014}.
\\\\
We consider a a double well potential in Jordan frame,
\begin{equation}
V_J= -\frac{1}{2}m^2\phi^2+\lambda \phi^4=F(\phi)^2V(\varphi)
\label{vj}
\end{equation}  
its Einstein frame action is given by
\begin{equation}
S_E=\int d^4x \sqrt{-g_E} \bigg(\frac{1}{2} R_E-\frac{1}{2}\partial^{\mu}\varphi\partial_{\mu} \varphi -V_E \bigg),\:\:V_E=\frac{V_J}{F^2}= -\frac{1}{2}m^2e^{-c \varphi}+\lambda
\label{sef}
\end{equation} 
where the potential $V_E$ matches the plateau potential of Eq.(\ref{v1}) if we set $\lambda=V_1$ and $\frac{1}{2}m^2=V_0$. The positive potential with $V_J=\frac{1}{2}m^2\phi^2+\lambda\phi^4$ is the simplest choice in many scalar-tensor modelling \cite{wetterich2014}.
\\\\
In the Ekpyrotic phase, the Einstein frame equation of state parameter is given by Eq.(\ref{ekpysol})
\be
\epsilon=\epsilon_E=\frac{c^2}{2} =\frac{2}{6-K}.
\label{epsilon}
\ee
Although our primary attention is in the Einstein frame dynamics, the Jordan frame dynamics, especially the scaling behaviour of the parameters play a critical role. The equation between $H_J$ and $H_E$ is given as 
\be
a=\sqrt{F}a_J;\:\:\:dt=\sqrt{F}dt_J;\:\:H_E=\frac{a_{,t}}{a}=\frac{1}{\sqrt{F}}\bigg(H_J+\frac{F_{,t_J}}{2F}\bigg), \: \dot{\phi}=\frac{d\phi}{dt}.
\ee
\be
H_J+ \frac{c}{2}\dot{\varphi}\phi=H_J+\frac{\dot{\phi}}{\phi}=H_E\phi.
\label{hjhe1}
\ee
\be
H_E=\pm \sqrt { \bigg(1-\frac{K}{6}\bigg)\frac{\dot{\phi}^2}{\phi^2}+\frac{1}{3\phi^4}\bigg(-\frac{1}{2}m^2\phi^2+\lambda\phi^4\bigg)}=\pm \sqrt{\frac{1}{6}\dot{\varphi}^2 -\frac{1}{6}m^2e^{-c\varphi} +\frac{1}{3}\lambda}.
\label{hjhe}
\ee
We note that from Eq.(\ref{hjhe}) that the "Pre-Ekpyrotic" phase of \cite{khoury}-\cite{khoury1} corresponds to a stable $3H_E^2\sim\lambda$  where the sum of the negative (mass) potential term and the kinetic term is small comparing to $\lambda$ 
\be
\frac{m^2}{2}e^{-c\varphi}\sim \frac{1}{2}\dot{\varphi}^2\ll \lambda.
\label{m2}
\ee
In Jordan frame, this corresponds to a nearly conformal coupling $K\simeq 6$ and a large $\phi^2$ ($\varphi>0$) for some small $m^2$. As $\varphi$ (and $\phi^2$) decreases, when the negative potential $-\frac{1}{2}m^2e^{-c\varphi}$ dominates over $\lambda$, one enters the "Ekpyrotic phase" .  
\\\\
At the beginning of the Ekpyrotic phase, starting with a large positive field ($\varphi>0$), a negative field speed ($\dot{\varphi}<0$) and large negative time $t$, the evolution of system in Einstein frame follows the work of \cite{khoury}-\cite{khoury1} with solution in Eq.(\ref{ekpysol}). From Eq.(\ref{epsilon}), the unnaturally large value $c\sim 10^{28}$ can be easily understood in Eq.(\ref{ck}) as $K$ having a near conformal coupling value. 
As $K$ moves away from the conformal coupling value to $6>K> 16/3$, we have $\epsilon> 3$ towards the end of Ekpyrotic phase. From Eq.(\ref{F}) we take the corresponding starting $\phi$ at $\phi>0$, with $\dot{\phi}=\frac{1}{2} (c\phi) \dot{\varphi}<0$. In this scenario the Jordan frame field $\phi$ rolls down from a large positive field to its bottom ($\phi=1$ in $M_p$ unit, corresponding to the Einstein frame pre-Ekpyrotic phase) and rolls up the potential towards the $\phi =0$ (corresponding to the Ekpyrotic and Kinetic energy dominant phase). At the end of Ekpyrotic phase, $K$ is still not far away from the conformal coupling value.  We note that scale invariant curvature spectrum is also obtained in \cite{li}-\cite{piao2} using Jordan frame model with nearly scale invariant potentials, though this spectrum are inflation related.
\\\\
After the Ekpyrotic phase, the scale factor continues to shrink in the Einstein frame (which is also true in Jordan frame for slow varying $\phi>0$). The Hubble constant increases towards the Planck scale. As the canonical kinetic term dominates over the potential with decreasing scale factor, the energy density which scales as $\sim a^{-6}$ can only continue to increase since none of the known cosmological components can slow it down. 
\\\\
To avoid reaching the Planck energy density, one way is to consider the existence of a high energy density fixed point below the Planck scale. Scalar-Tensor theories in the form of Eq. (\ref{s1}) are used to describe the universe's evolution from very high energy densities to low energy densities over a large energy density scales. Conformally invariant fixed point $K= 6$ and $m^2= 0$ provides an important solution. In \cite{wetterich2}, for a positive potential 
\be
V_J(\phi)=+\frac{1}{2}m^2\phi^2+\lambda\phi^4,
\label{sp}
\ee
functional renormalisation group flow calculation is done for the parameter $y$ where $\frac{1}{y^2}= \frac{m^2}{\phi^2}$, where $m^2$ plays the role of  the renormalisation energy scale $ k^2$, here the direct proportionality of $m^2$ to $k^2$ is assumed.  For $m \rightarrow 0$ much faster than $\phi \rightarrow 0$ ($y\rightarrow \infty$), the system will approach the conformally invariant (infra-red) fixed point described above. This infra-red fixed point can be identified to the Einstein-Hilbert action in Einstein frame. The UV limit is reached by $y\rightarrow 0$, which is the small $\phi$ region where $\phi^2\ll m^2$, which is  the UV fixed point of the asymptotically safe quantum gravity \cite{wetterich2}-\cite{litim2004}.
\\\\
To avoid the Hubble radius shrinks to the Planck scale, another way is to consider, due to double well potential, before the system reaches the Planck energy density, it could go through a phase transition at the conformal fixed point (where $m^2=0$, $K=6$, $H_E=0$) into the symmetric phase with potential described by Eq.(\ref{sp}). The idea of the scalar inflaton having both a symmetric and a broken symmertic phase is not new. A second order phase transition in quantum gravity around some critical field value $\phi^2_c$ has also been proposed in \cite{falls}.
Early inflation scenario proposed Guth \cite{guth2} assumes a first order phase transition from the symmetric phase to the broken symmetric phase. The New inflation scenario  \cite{linde2} assumes a continuous phase transition also from the symmetric phase to the broken symmetric phase.
In a flat space $R_J=0$ scenario, this is the most common mechanism used for phase transition studies \cite{luscher}. In cosmological evolution, as the energy scale varies across a large range that a scalar field with a given symmetry is expected to go through phase transitions, which could lead to inflation \cite{guth2} as well as in the production of cosmological defects \cite{kibble}. 
\\\\
In this work, we shall follow \cite{steinhardt} in working in  the Einstein frame for $H_E$ ($\dot{H}_E$) evolution and uses the field ($\varphi$) roll in the Einstein potential to consider the perturbation power spectrum in inflation and contraction. However, we shall use the Jordan frame field ($\phi$) potential to follow its phase transition behaviour.
\\\\
More specifically, we consider the scenario that as the energy density increases the Jordan frame system enters into the scaling (Ginsberg) region of a phase transition, in which system parameters follow scaling laws. Its broken symmetric phase (renormalised) mass term $-m^2\phi^2<0$ in Eq.(\ref{vj}) will go through zero into the symmetric phase scaling region with $+m^2\phi^2>0$. If the phase transition fixed point is effectively at the conformal invariant fixed point, as the renormalised mass goes through zero, the coupling $K$ approaches the conformal coupling limit   $K\rightarrow 6$ and then moves away. The end result is that in Einstein frame, the mass term $m^2(\rho)$ (and $\lambda(\rho)$) also have an energy density dependence, with a deriative w.r.t  to $t$ and $\rho$ given by
\be
m^2_{,t}= m^2_{,\rho} \dot{\rho}(a).
\ee
When $\rho$ increases over time, we have $\dot{\rho}>0$ and the renormalised mass flow is $m^2_{,\rho}<0$.  As $c^2\propto (6-K)^{-1}$, the corresponding rate of change for $c$ is $c_{,\rho}>0$ towards the critical point. 
\subsection{Scaling region in the Broken symmetric phase}
From the above discussion, in the scaling region of the broken symmetric phase, we write that the mass term is dependent on the energy density of the Einstein frame $\rho$, as
\be
V_J= -\frac{1}{2} m^2(\rho)\phi^2+\lambda\phi^4
\ee
so that 
\be
3H_E^2= \frac{1}{2}\dot{\varphi}^2-\frac{1}{2}m^2(\rho) e^{-c\varphi}+\lambda.
\label{he2}
\ee
Keeping in mind that it is the parameters in Jordan frame that scale as energy density increases, we differentiate Eq. (\ref{he2}) w.r.t the time $t$ in Einstein frame and obtain
\be 
3\frac{d}{dt} (H_E^2)= \dot{\varphi} \ddot{\varphi}+V_{,\varphi} \dot{\varphi}-\frac{1}{2 }\frac{m^2_{,t}}{\phi^2} -\frac{1}{2}\frac{m^2}{\phi^2} \bigg(-\dot{c}\varphi\bigg) +\dot{\lambda}
\label{dh2}
\ee
\be
\dot{H}_E =-\frac{1}{2}\dot{\varphi}^2-\frac{1}{12H_E} \bigg[\frac{m^2}{\phi^2}\bigg(\frac{m^2_{,t}}{m^2} -\dot{c}\varphi\bigg) -2\dot{\lambda}\bigg].
\label{dotH1}
\ee
We note that IF the mass parameter $m^2$, $c$ and $\lambda$ are constant, we will recover the usual result $\dot{H}_E=-\frac{1}{2} \dot{\varphi}^2$ and there is no built-in mechanism to switch off the shrinking of the scale factor.  
\\\\
Writing explicitly that
\be
m^2_{,\rho}=\frac{dm^2}{d\rho},\:\:\:c_{, \rho}=\frac{d c}{d\rho}, \:\:\: \lambda_{, \rho}=\frac{d\lambda}{d\rho}.
\label{dotc}
\ee
where respectively $m^2_{,\rho}$ and $\lambda_{,\rho}$ are the renormalised mass flow and renormalised coupling flow due to energy density increases, while $c_{, \rho}$ is a result of the non-conformal coupling flow towards the conformal value. 
 We can rewrite Eq.(\ref{dotH1}) as
\be
\dot{H_E}=-\frac{1}{2}\dot{\varphi}^2 -\frac{\dot{\rho}}{12H_E}\bigg[\bigg(\frac{m^2_{,\rho}}{m^2}-c_{, \rho} \varphi\bigg) \frac{m^2}{\phi^2}-2\lambda_{, \rho}\bigg]
\label{dotrho1}
\ee
We assume that scaling laws here is qualitatively similar to the flat space canonical $\lambda \phi^4$ theory \cite{luscher}, we have
\be
m^2=a_0 (\rho_c-\rho) ^l;\:\;\:\:\;\:\lambda=-\frac{\lambda_0} {\ln(\rho_c-\rho)}.
\label{scaling}
\ee
where $a_0>0$ and $\lambda_0>0$ are constants, and the exponent $l>0$ describes the effect of mass renormalisation in the scaling region. (We note that the mean field value for $l$ is $l_{MF}=1$ for the flat space $\lambda\phi^4$ theory in 4-dimensions. A High Temperature series analysis of the $\lambda \phi^4$ theory produces $l>l_{MF}$ \cite{vladikas}. It is also shown that $\lambda$ goes to zero logarthimically. 
\\\\
The important feature of these scaling laws is that as $\rho$ approaches $\rho_c$, initially the mass term, which dominates the potential, scales to zero faster than the $\lambda$ term. Subsequently, the $\lambda$ term will dominate the potential but will vanish logarithimically at the phase transition point.
\\\\
In Einstein frame, for the potential $V_E$ to rise as energy density increases towards the phase transition, the mass term $m^2/\phi^2 =m^2e^{-c\varphi}$ will need to reduce towards zero. Also at finite $\varphi<0$ one has $c\rightarrow \infty$ towards the phase transition point. A consistent choice for $\phi$ and $c$ will be
\be
\phi^2=b_0(\rho_c-\rho)^{q},\:\:\:\:\;\:c=\frac{1}{\varphi}\ln \phi^2 \propto q\ln(\rho_c-\rho).
\ee
This choice means that  as $m^2(\rho)\rightarrow 0$, the field $\phi$ rolls to zero, which is the location of the central peak of the broken symmetric potential and the minimum of the symmetric potential. This field roll corresponds to $c$ goes to its conformal limit. The renormalisation of field $\phi$ value near phase transition plays no effective role here.
\\\\
For the exponent $l>q>0$, as $\rho\rightarrow \rho_c$ from below, the mass term
\be
\frac{m^2}{\phi^2}=\frac{a_0}{b_0}(\rho_c-\rho)^{(l-q)}\rightarrow 0.
\label{m2phi2}
\ee
Also we have
\be
\frac{m^2_{,\rho}}{m^2}= \frac{-l}{ (\rho_c-\rho)};\:\;\:\frac{2\phi_{,\rho}}{\phi}=\frac{-q}{(\rho_c-\rho)};\:\:\: \lambda_{, \rho}=\frac{\lambda}{ (\rho_c-\rho) \ln(\rho_c-\rho) }.
\label{scalingrelation}
\ee
As $c$ diverges logarthimically, we have near the critical point
\be
c_{, \rho}\varphi= 2\frac{\phi_{,\rho}} {\phi}.
\ee
We note that the energy density of the patch with scale factor $a(t)$ is $\rho(a)=a^{-2\epsilon}$. Its time derivative is given by
\be
\dot{\rho} = \dot{\epsilon}\frac{d}{d \epsilon} a^{-2\epsilon} =-2\epsilon\rho H_E-2\dot{\epsilon}\rho \ln \bigg(\frac{a}{a_p}\bigg),
\label{dotrho}
\ee
where $a_p$ is some reference scale that $a>a_p$.
The equation of state parameter $\epsilon$ from Eq.(\ref{eos}) in Einstein Frame is given by in the broken symmetric phase 
\be
\epsilon =\frac{3}{2}\bigg(1+\frac{P}{\rho}\bigg)=3\frac{ \dot{\varphi}^2}{\dot{\varphi}^2+2V(\varphi)}= 3\frac{ \dot{\varphi}^2}{\dot{\varphi}^2-\frac{m^2}{\phi^2}+2\lambda}.
\label{eos2}
\ee
Starting with large field speeds and a decreasing mass term $m^2/\phi^2$, in the scaling region where $\epsilon$ remains nearly constant such that $\dot{\epsilon} \sim 0$. From Eq.( \ref{dotrho1}) we obtain
\be
\dot{H}_E=-\frac{1}{2}\dot{\varphi}^2+\frac{1}{6}\frac{\epsilon \rho}{(\rho_c-\rho)} \bigg((q-l)\frac{m^2}{\phi^2}-\frac{2\lambda}{\ln(\rho_c-\rho)}\bigg)=\frac{1}{2}\dot{\varphi}^2\bigg[-1+\frac{1}{\rho_c-\rho}\bigg((q-l)\frac{m^2}{\phi^2}-\frac{2\lambda}{\ln(\rho_c-\rho)}\bigg)\bigg].
\label{dotH}
\ee
In this case, in the scaling region far away from the critical point, the $m^2/\phi^2$ term dominates over the $\lambda$ term, $\dot{H_E}$ remains negative. Moving towards the phase transition, one will reach the region where the $\lambda$ term with the logarithmic factor in Eq.(\ref{dotH}) starts to dominate, (taking initially a dominant field speed $\dot{\varphi}^2\sim \rho$,) here $\dot{H}_E$ becomes positive and grows rapidly as $\rho \rightarrow \rho_c$. The field speed will reduce rapidly to close to zero and  from Eq.(\ref{he2}), the negative $V_E$ allows the value $H_E=0$ to be reached continuously.
\subsection{Scaling region in the Symmetric phase}
Kibble \cite{kibble}-\cite{kibble2} considers that in a large physical system going through a second order phase transition, the correlation length does not have enough time to reach infinity before the phase transition is completed. In this scenario, phase transition occurs when the mass term $-\frac{1}{2} m^2\varphi^2$ is frozen and simply flips to $+\frac{1}{2}m^2\varphi^2$ after a finite time. For a small Hubble radius this may be problematic. If the cut-off scale factor $a_p$ occurs at Planck scale ($\sim (10^{19} GeV)^4$), a phase transition at energy scale $\sim (10^{14}GeV)^4$ may involve finite size effects or modification of scaling laws. For our purpose,   we assume that the modification of scaling exponents does not affect our result. We also note that when the Hubble radius becomes nearly infinite ($H_E=0$) around the phase transition point, the Kibble mechanism is perfectly valid around this region. 
\\\\
In the symmetric phase, the Einstein frame Hubble constant turns positive, $H_E>0$. We consider a scenario that for a short time the energy density can continue to grow such that $\dot{\rho}>0$. In the symmetric phase the potential is given by, for $\varphi<0$
\be
V(\varphi)=\frac{1}{2}m^2(\rho)e^{-c\varphi}+\lambda
\label{vsym}
\ee
with Hubble parameter
\be
3H_E^2=\rho(a)=\frac{1}{2}\dot{\varphi}^2+\frac{1}{2}m^2e^{-c\varphi}+\lambda;\:\;\: \varphi<0.
\label {H2}
\ee
If the model is in flat space, the scaling laws on both sides of the phase transition are similar. We shall assume that the scaling laws remain similar in our interested region on both sides of the phase transition. We have $m^2$ and $\lambda$ scales as
\be
m^2\propto a_1(\rho-\rho_c)^l;\:\:\:\: \lambda =-\frac{\lambda_1}{ \ln(\rho-\rho_c)};\:\: l>0.
\label{symsls}
\ee
Assume also that the phase transition is also a conformal fixed point for the symmetric phase, we have
\be
c=\frac{1}{\varphi}\ln \phi^2,\:\:\:\phi^2\propto (\rho-\rho_c)^q,
\label{cnphi}
\ee
and also
\be
\dot{H}_E=-\frac{1}{2}\dot{\varphi}^2 +\frac{\dot{\rho}}{12H_E}\bigg[\bigg(\frac{m^2_{,\rho}}{m^2}-\frac{2\dot{\phi}}{\phi}\bigg) \frac{m^2}{\phi^2} -2\lambda_{, \rho}\bigg].
\label{dotrho2}
\ee
From Eq.(\ref{symsls}) and Eq.(\ref{cnphi}), we obtain
\be
\frac{m^2_{,\rho}}{m^2}=\frac{l}{\rho-\rho_c},\:\:\frac{2\dot{\phi}}{\phi}=\frac{q}{\rho-\rho_c},\:\:\lambda_{,\rho}=\frac{\lambda}{(\rho-\rho_c)\ln (\rho-\rho_c)}.
\ee
and
\be
\dot{H}_E=-\frac{1}{2}\dot{\varphi}^2 + \frac{\dot{\rho}}{12H_E} \frac{1}{(\rho-\rho_c)}\bigg((l-q)\frac{m^2}{\phi^2}-\frac{2\lambda}{\ln(\rho-\rho_c)}\bigg).
\label{dothe2}
\ee
We see that for $H_E>0$ and $\dot{\rho}>0$, $\dot{H}_E>0$ is maintained for a very small $\dot{\varphi}^2$.
\subsection{The bounce}
From the Friedmann equation $3H_E^2=\frac{1}{2}\dot{\varphi}^2-\frac{1}{2}m^2e^{-c\varphi}+\lambda$, we see that for non-zero field speed $\dot{\varphi}>0$, $H_E=0$ can in principle be reached in the broken symmetric phase, where the equation of state parameter $\epsilon \rightarrow \infty$. However, the mass term $-m^2/\phi^2$ freezes and changes sign, the $H_E=0$ will not be reached but $H_E<0$ simply changes to $H_E>0$ and implements the bounce. Within a finite size Hubble radius, the asymptotic limit $\lambda$ will be finite.
\\\\
We consider the scenario that $H_E=0$ can be reached and the mass term in Eq.(\ref{eos2}) can move continuously from negative value towards positive value.  From Eq.(\ref{dotrho}), we have initially $H_E=0_{+} $
\be
\dot{\rho}=-2\epsilon \rho \bigg (H_E+\frac{\dot{\epsilon}}{\epsilon}\ln\bigg(\frac{a}{a_p}\bigg)\bigg)>0.
\label{dotrhohe}
\ee
Here, the energy density can continue to increase $\dot{\rho}>0$ if $\dot{\epsilon}<0$ and $a>a_p$. Here $a_p$ plays the role of a "minimum scale factor". $\dot{\epsilon}<0$ is possible  since the mass term increases in Eq.(\ref{eos2}) will take the correspoding value of $\epsilon$ to come down from infinity to $\epsilon<3$ as $H_E$ passes through $H_E=0$ to $H_E>0$. The system reaches the symmetric phase where the mass term in the potential $V_E$ takes positive value. This energy density continues to grow until
\be
H_E=-\frac{\dot{\epsilon}}{\epsilon} \ln \bigg(\frac{a}{a_p}\bigg),
\ee
where this $H_E>0$ system reaches a peak $\rho$ value with $\dot{\rho}=0$ ($H_E\geq\sqrt{\rho/3}$) and $\dot{H}_E<0$ in the symmetric phase. The bounce from this point leads naturally to an expanding phase. When $\dot{\rho}=0$, if the potential takes the value $V_E<\frac{1}{2}\dot{\varphi}^2$,  we have $\epsilon>2$ and the system enters the radiation epoch directly. If $V_E>\dot{\varphi}^2$, we have $\epsilon<1$ and the system will go through a short inflationary epoch. 
\\\\
We should point out from the results in Wetterich \cite{wetterich3} that if the energy density after the bounce is large enough such that $V_E\gg \dot{\varphi}^2$, the system possesses a global solution which initially has a long inflationary phase and eventually crossovers and enters into a late dynamical dark energy epoch.
\subsection{After the bounce}
After the bounce, the field speed relation between Einstein frame and Jordan frame is given by
\be
\dot{\varphi} =\frac{2}{c}\frac{\dot{\phi} }{\phi}+\frac{\dot{c}}{c}\varphi.
\ee
Consider the point where $-\varphi\sim O(1)$. We take that $c$ remains positive, and that the Jordan frame field $\phi$ lands at $0<\phi<1$ in the symmetric potential 
\be
V_J(\phi)=\frac{1}{2} m^2(\rho)\phi^2+\lambda\phi^4.
\label{jfpotl}
\ee
For $\phi$ to roll down the symmetric potential, we need the kinetic energy term to have the right sign $K<0$ (and $c^2<1/3$), one has $\frac{\dot{c}}{c}\ll1$, the negative Einstein frame field speed $\dot{\varphi}<0$ corresponds to $\dot{\phi}<0$ (and $\frac{d\phi}{dt_J}<0$), so that $\varphi$ will continue its roll to the $\varphi \rightarrow -\infty$ while $\phi$ rolls towards $\phi\sim 0$.
\\\\
After $\dot{\rho}=0$, for an potential $V_E$ that is slightly larger than $\rho_c$, which is stipulated by KS, after the bounce, the system goes into either a short inflation or the radiation dominant phase. For bounce universe models considered in the literature (ref), the Einstein frame potential is
\be
V_E(\varphi)=\frac{1}{2}m^2(\rho) e^{-c\varphi}+\lambda,
\label{pospotl}
\ee
where small $m^2$ remains fixed from the beginning. The potential will reduce to $\lambda$ asymptotically when $c$ moves away from infinity and $\varphi \rightarrow -\infty$, where the energy density $\rho\rightarrow 0$ is assumed.
\\\\
Here we can assume explicitly that the $m^2$ term also has an energy density dependence, so that as $\rho\rightarrow 0$, we have also $m^2(\rho) \rightarrow \rho^l$ ($l>0$) in addition to $e^{-c\varphi}=\phi^{-2} \rightarrow \rho^{q} $, such that the term $\frac{1}{2}m^2e^{-c\varphi}$ reduces more rapidly in the low energy density limit. $\lambda$ continues to play the role of the cosmological constant.
\\\\
We recall that a conformal infra-red fixed point for the (Einstein frame) positive potential in Eq.(\ref{pospotl}) is found in \cite{wetterich2}. That a non-minimal coupling model with $K\sim 6$ at late time universe has recently found observational support \cite{hrycyna}. It is possible that as $\rho\rightarrow 0$, the Jordan frame system parameters flow into a conformal infra-red fixed point according to some scaling laws similar to that in the symmetric phase above. 
The field acceleration equation
\be
\ddot{\varphi}= -3H_E\dot{\varphi}-V_{,\varphi}=-3H_E\dot{\varphi}+\frac{1}{2}cm^2e^{-c\varphi}
\ee
shows that both terms on the R.H.S. provide friction on the field speed $\dot{\varphi}<0$. The negative field speed $\dot{\varphi}<0$ will eventually reach $\dot{\varphi}>0$. We wish to consider whether the system can return to the starting point of Khoury-Steinhardt \cite{khoury}-\cite{khoury1} after a quick discussion on the cosmological constant.
\subsection{Implication for the cosmological constant}
Here, we come to a convenient place to consider the issue of cosmological constant value.
 In Jordan frame, the Einstein frame cosmological constant $\lambda$ becomes the four point coupling constant of a $\phi^4$ theory. 
\\\\
There are attempts to use the physics of the renormalised $\lambda \phi^4$ theory to address the small cosmological constant problem.
Polyakov in \cite{polyakov} considers the conformally flat universe where $g_{\mu \nu}=\phi^2\delta_{\mu \nu}$.
In Euclidean time, the Einstein-Hilbert action becomes
\be
S =\int d^4 x \bigg(\frac{1}{2} (\partial \phi)^2+\lambda\phi^4\bigg). 
\ee 
From Eq.(\ref{eos}) it can be argued that in the infra-red limit ($\rho\sim 0$) and large $\phi$ limit, $\lambda$ will be screened to zero.  However, once the curvature $R\sim \phi^{-3}\partial^2\phi$ is allowed, the screening could fail for small $\phi$. 
\\\\
In \cite{wetterich2}, the authors use the functional renormalisation group to consider the Jordan frame symmetric potential without the $\lambda \phi^4$ term. They find that in the infra-red regime and large $\phi$ region, the Einstein frame cosmological constant is the effective potential $V_E \sim \frac{m_g^2}{\phi^2}$, where $m_g$ is the scalar graviton mass and $\phi=M_{pl}$.
\\\\
The value of the dimensionless cosmological constant is estimated by Barrow and Shaw \cite{barrow} to be
\be
\lambda =1.7\times 10^{-122}.
\ee
Since $\lambda$ is the measured value at low energies, Hsu and Zee \cite{zee} postulates that given the Planck mass $M_{pl}$ and the mass scale of the observable universe $M_U=2\times 10^{-33}eV$, which matches the smallest upper bound of the massive graviton, de Rham et al. \cite{deRham}. The vacuum energy density takes the value
\be
\lambda \phi^4\sim M_{pl}^2M_U^2 \sim \bigg(10^{19}GeV \times 2\times 10^{-33} eV\bigg)^2 \sim \bigg(4.5\times 10^{-3}eV\bigg)^4,
\ee
which provides an order of magnitude match for the observed vacuum energy density. If we use $\phi^2=M_{pl}^2$, we will obtain
\be
\lambda=\frac{M_U^2}{M_{pl}^2} \sim 10^{-122}.
\ee
Our approach is slightly different. Here, the Jordan frame potential is in the symmetric phase at low energy density, if the renormalised scalar graviton mass term $m_g^2/\phi^2$ is much larger than $\lambda$, the effective cosmological constant term is 
\be
\lambda =V(\varphi) \sim \frac{1}{2} \frac{m^2}{\phi^2}.
\ee  
If the renormalised mass term is much smaller than $\lambda$, as discussed in \cite{luscher}, if we assume that the $\lambda\phi^4$ system is sitting in the scaling region of a phase transition to a broken symmetric phase, the renormalised  coupling constant $\lambda$ takes on the same value as its counterpart in the scaling region of broken symmetric phase \cite{luscher}, such that, in terms of the definition in \cite{luscher}
\be
\lambda =3\frac{m_g^2}{\phi^2}.
\ee 
We take the scalar graviton mass to be close to $10^{-33} eV$ in \cite{deRham}, so that we have an estimate of the cosmological constant as
\be
\lambda \sim 3\times \bigg(\frac{10^{-33}eV}{10^{28}eV}\bigg)^2 =3\times 10^{-122}.
\ee
%
%
\subsection{Transition back to broken symmetric phase}
Next, we consider a situation in the Jordan frame symmetric potential that after long enough time, $\phi$ goes back to a small $\phi\sim 0$ and a phase transition to the broken symmetric phase occurs near the conformal point. From Eq. (50), we further assume that near the phase transition we have $\dot{\phi}\sim 0$ ($\dot{\varphi}\sim 0$) and $\dot{c}\sim 0$ (although $c\gg1$). 
\\\\
Before the phase transition, a field $\varphi$ rolling down the plateau potential $V=+\frac{1}{2}m^2 e^{-c\varphi}$ is known to admit an inflation phase with a scale invariant power spectrum \cite{lyth}. Immediately after phase transition, from \cite{lyth} with $t>0$, we continue to have (at small $m^2$) a slow roll inflationary phase in the broken symmetric potential $V_E=-\frac{1}{2}m^2e^{-c\varphi}+\lambda$, with $c$ moves away from $\infty$ and  $\varphi \rightarrow 0$ in the negative $\varphi$ ($\varphi<0$) region.
\\\\
Initially, $e^{-c\varphi}$ is large, but it can be sufficiently suppressed by the small $m^2$ term. As $c\varphi$ moves away from $-\infty$, $\phi^2=e^{c\varphi}>0$ rolls away from the conformal phase transition point at $\phi\sim 0$. Consequently, the renormalised mass squared $m^2$ also moves away from its small value and will increase to a larger but nearly fixed value $M^2$ outside the scaling region. This large mass $M$ no longer suppresses the $e^{-c\varphi}$ term.  The negative potential $-\frac{1}{2}M^2e^{-c\varphi}$ takes on a large negative value which leads to a fast $\varphi$ field roll ($\dot{\varphi}>0$) down the negative potential. As the field speed converts to radiation and matter density (reheats) and $e^{-c\varphi} \rightarrow O(1)$ is reached, the Hubble constant $H_E$ is dominated by radiation and matter density. Field roll in this section of the model potential provides a similar scenario to the standard inflation and hot big bang model except that in this case a slow roll inflationary phase already occurred in the symmetric phase before the phase transition. A larger $m^2$ that does not full suppress the $e^{-c\varphi}$ term will allow for a shorter inflationary phase after phase transition. Most importantly, a long inflationary era in the late symmetric phase could provide the sought after "a large patch of smooth background Initial condition" of the inflationary paradigm. The graceful exit comes from the renormalised mass increase due to the system moving away from the scaling region.
\\\\
Lastly, we consider the large time and large $\varphi>0$ asymptotic solution. This large $\varphi$ limit is the conformal fixed point which Khoury Steinhardt \cite{khoury}-\cite{khoury1} starts its pre-Ekpyrotic roll. One question is whether this same initial condition can be reached again. The field acceleration equation, for $\dot{\varphi}>0$ is
\be
\ddot{\varphi}=-3H_E \dot{\varphi}-\frac{1}{2} cm^2e^{-c\varphi} <0,
\label{dece}
\ee
this acceleration slows positive $\varphi$ field rolls and allows for $\dot{\varphi}\rightarrow 0$. We assume when approaching the large field conformal fixed point, similar scaling laws as above holds for the renormalised mass and the field $\phi$
\be
m^2(\rho) =m_0^2\rho^l; \:\:\:\:\phi^2=\phi^2_0\rho^{-q},\:\:\:\frac{m^2}{\phi^2}=\frac{m_0^2}{\phi_0^2} \rho ^{l+q}.
\label{slaws}
\ee
where $m_0$ and $\phi_0$ are constant. 
\be
c\varphi=\ln \phi^2;\:\:\:c=\frac{1}{\varphi}( \ln \phi_0^2-q\ln \rho),\:\:\:\:\varphi\frac{dc}{d\rho}=\frac{-q}{\rho}.
\ee
We could continue to assume that $\lambda(\rho)$ reduces logarithimically as $\rho$ reduces such that
\be
\lambda =-\frac{\lambda_0}{ \ln (\rho)}.
\ee
the R.H.S of Eq.(\ref{dece}) remains negative and in the small $\dot{\varphi}$ region, we have
\be
\ddot{\varphi} \sim - \frac{q}{\varphi} (\ln \rho) \rho ^{l+q}.
\label{ddotvarphi2}
\ee
We consider the case that the small field speed is "reversed" ($\dot{\varphi}<0$) at some small $\rho$ value. Since $\ddot{\varphi}$ is small in the small $\rho$ region, the field speed $\dot{\varphi}^2$ remains small.
\\\\
As $\rho\rightarrow 0$, the Hubble constant approaches firstly
\be
H_E=+\frac{1}{\sqrt{3}}\sqrt{\frac{1}{2}\dot{\varphi}^2-\frac{1}{2}m^2e^{-c\varphi}+\lambda} \sim\frac{1}{\sqrt{3}}\sqrt{\frac{1}{2}\dot{\varphi}^2+\lambda}.
\label{Helf}
\ee
We recall that from Eq.(\ref{dotrho1}), as the $m^2/\phi^2 \rightarrow 0$
\be
\dot{H_E}=-\frac{1}{2}\dot{\varphi}^2 -\frac{\dot{\rho}}{12H_E}\bigg[\bigg(\frac{m^2_{,\rho}}{m^2}-c_{, \rho} \varphi\bigg) \frac{m^2}{\phi^2}-2\lambda_{, \rho}\bigg]=-\frac{1}{2}\dot{\varphi}^2 +\frac{1}{6}\frac{\dot{\rho}}{H_E} \lambda_{, \rho}.
\label{dotrho2}
\ee
If $\lambda_{,\rho} >0$, as $\rho\rightarrow 0$, we have $\dot{\rho}<0$, the last term in Eq.(\ref{dotrho2}) is negative,
so that $\dot{H}_E<0$. Also from Eq.(\ref{slaws}) and Eq.(\ref{Helf}),  we have $\lambda \rightarrow 0$ and if $H_E \rightarrow \frac{1}{2}\dot{\varphi}^2$ asymptotically. This is close to the conformal fixed point at large $\varphi$.
\\\\ 
Since we have $\dot{\varphi}^2>0$ in the (large field) infra-red limit, the energy density has a minimum value such that $\rho\rightarrow \rho_{min}$, ($\rho_{min}>0$) and at the region where $\dot{\rho} \rightarrow 0$, we have from Eq.(\ref{eos2})
\be
H_E=-\frac{\dot{\rho}}{\rho}-\frac{\dot{\epsilon}}{\epsilon} \ln \bigg(\frac{a}{a_p}\bigg),\:\:\: a\gg a_p.
\ee
We recall that from  Eq.(\ref{ekpysol}), we have $\epsilon =\frac{c^2}{2}$ and the conformal fixed point corresponds to $c\rightarrow \infty$, so that
\be
\frac{\dot{\epsilon}}{\epsilon}=2\frac{\dot{c}}{c}= \frac{\dot{\rho}}{\rho}\frac{2}{ \ln \rho}
\ee
\be
H_E =-\frac{\dot{\rho}}{\rho}\bigg(1+\frac{2}{\ln \rho}\ln\bigg(\frac{a}{a_p}\bigg)\bigg)
\label{HERB}
\ee
For some small $\rho$, from Eq,(\ref{HERB}) $H_E$ will become  negative.  The end result is that $H_E$ takes on a negative value 
\be
H_E \sim -\frac{1}{\sqrt{3}}\sqrt{\frac{1}{2}\dot{\varphi}^2-\frac{1}{2}m^2 e^{-c\varphi}+\lambda}.
\ee
We now enter a contracting universe $H_E<0$ and $\dot{H}_E<0$, and the energy density will start increasing with $\dot{\rho}>0$. In this case, we recover the initial condition for the pre-ekpyrotic phase of KS \cite{khoury}-\cite{khoury1}, however one still has to switch to $t<0$ to start the contraction process. One obtains a cyclical model as postulated by Ijja and Steinhardt \cite{steinhardt2}
\subsection {A Conformal Cyclic Cosmology scenario}
Apart from a bounce from an expanding phase to a contracting phase, after $\dot{\varphi}<0$, the potential $V_E=-\frac{1}{2}m^2e^{-c\varphi}+\lambda$ could transition back to $V_E=+\frac{1}{2}m^2 e^{-c\varphi}+\lambda$ in the small $\rho$ region following the Kibble argument \cite{kibble}-\cite{kibble2}. In Jordan frame, the field $\phi$ in the symmetric phase starts a slow field roll back towards $\phi\sim 0$ region. Following this pathway, one could recover a cyclical model without a contracting phase, which shares some features with the conformal cyclic cosmology (CCC) proposed by Penrose \cite{penrose}. Here the conformal factor $\Omega$ in CCC is identified to $\phi$ in our model as ($\Omega=\phi^2$). In CCC, the universe in far future goes into a phase which is inflationary and cold where the space-time metric is dominated by the conformal metric. Here we identify this phase in CCC with the conformal fixed point of the Jordan frame system Eq.(\ref{s1}) with potential in Eq. (\ref{sp}). In our model, to reach the conformal fixed point, both the known massive particles density and the mass of the scalar field (which identifies with the inflaton in an inflationary paradigm) need to be negligible. The non-zero mass of the scalar inflaton arises for phenomenological reasons. Both models postulate the existence of a mechanism which connects the cold phase of a space-time region (aeon) to a hot phase which should describe the observed early universe.  CCC postulates that where the universe loses all scales, the cold conformal universe can be matched mathematically to a hot universe. Our model here relies on the conventional phase transition theory and after phase transition, the inflationary phase in the symmetric phase is connected to  a hot phase only when the mass term is large enough such that it does not suppress the field term $e^{-c\varphi}$. 
\subsection{Summary and Discussion}
After Planck (2013), single field plateau potentials beome the simplest choices to provide the inflationary scenario. From a lifted plateau potential, Khoury and Steinhard (KS) argue for an Ekpyrotic solution which could provide a reasonable curvature perturbation spectrum. This is a new paradigm and the challenge is in finding a non-singular bounce mechanism to bring the contracting universe back to the expanding universe. We take the KS potential in Einstein frame to a double well potential in Jordan frame and identify their model constants $V_0$, $V_1$ with the quartic coupling and mass squared, with $c^{-2}=(K-6)$ measures the deviation of the non-conformal coupling $K$ from its conformal limit. The initial condition for the KS pre-Ekpyrotic phase can be idenfied with the neighbourhood of a conformal fixed point in the Jordan frame system. 
\\\\
By postulating a phase transition at the conformal point at some critical energy density for the Jordan frame $\lambda \phi^4$ field, we use the known flat space scaling laws as approximation for the system parameters around the phase transition. We obtain a non-singular bounce solution which takes the system into its symmetric phase and to an expanding Hubble radius. This is the primary result. 
\\\\
In Einstein frame, after the bounce, the scalar field in this expanding universe is expected to roll down a positive plateau potential $V_E=\frac{1}{2}\frac{m2}{\phi^2} +\lambda$ and reaches the value of quartic coupling $\lambda$ asymptotically. If the system is in a phase where the mass term still dominates, one would obtain the correct order of magnitude estimate of the cosmological constant. If the system is in a phase where $\lambda$ dominates, if the symmetric system is inside the scaling region of a phase transition (which leads to a broken symmetric phase) at small field $\phi$, $\lambda$ matches the value of its broken symmetric phase counterpart which gives the correct present value estimate for the cosmological constant.
\\\\
We consider further that at low energy density inside the Hubble radius, the symmetric phase system admits (yet again) a phase transition back to the broken symmetric phase. The first thing to note is that at the end of the symmetric phase, the slow field roll in Einstein frame potential will lead to a long inflationary epoch. After a smooth phase transition, the field roll in Einstein frame potential can continue to produce inflationary phase. The inflationary phase in the symmetric phase before the phase transition can provide the large and smooth background initial condition that an inflation model requires to avoid large and uncontrolled fluctuations.
\\\\
We see that once the field $\phi$ rolls to value far away from the conformal point, the renormalised mass moves away from its scaling region and increases. The Einstein frame negative potential becomes large (and negative), the field speed $\dot{\varphi}$ from rolling down the potential increases and can be converted to radiation and matter density. Since $c$ reduces rapidly once the system moves away from the conformal point and when $\varphi$ moves across $\varphi=0$ into the positive value region, the negative potential shrinks rapidly in magnitude, the Hubble parameter is dominated by radiation and matter density and the standard big bang scenario is ensued. At very large time (large $\varphi$) and low energy density, the field speed $\dot{\varphi}$ reduces to very small value. We consider the scenario that the field speed decelerate enough to reach a negative field speed in the small field speed region. As energy density reduces, the rate of change in the equation of state parameter $\epsilon$ will lead to a bounce from an expansion phase $H_E>0$ to a contraction phase $H_E<0$, while $\dot{H}_E<0$ remains. This can be identified with the initial pre-Ekpyrotic phase condition postulated by KS . At small energy density, the Jordan frame system can also transit back to the symmetric phase at large field, the large $\phi$ will slow roll back to $\phi \sim 0$ and restart the process. This pathway has no contracting phase and phenomenologically resembles the conformal cyclic cosmology proposed by Penrose.
\\\\
This model possesses both an inflationary and an ekpyrotic solution, either of which could describe the observed history of our universe. However, the field roll in the symmetric phase at low energy density seems to be a critical component of any realisic model. This model could support a cyclical universe with a bounce phase proposed by Ijja and Steinhardt. It can also support a cyclical universe with no contracting phase which is phenomenologically similar to the CCC proposed by Penrose. The origin of the universe is therefore pushed back to the first cycles. In our model, there is a minimum scale factor $a_p$. In this case, a consistent postulate is that initially the system with a minimum scale factor $a=a_p$ sits at a conformally invariant phase transition point, with the effective field $\phi=0$, $\dot{\phi}=0$, $K=6$, $m^2=\lambda=0$ ($\rho=0$). The field $\phi$ dynamics near a conformal point can be sensitive to other plausible quantum effects such as the Coleman-Weinberg mechanism and will not be pursued here.
\section*{Acknowledgement}
The auther would like to thank Alan Vikman for helpful comments and Roger Penrose for an insightful explanation of the Conformal Cyclic Cosmology.

\end{document}